\documentclass[aps,pre,reprint,floatfix,superscriptaddress]{revtex4-2}

\usepackage{graphicx}
\usepackage{hyperref}
\usepackage{amsmath,amssymb,amsfonts}
\graphicspath{{figs/}}
\begin{document}

\title{Electrical conductivity of nanorod-based transparent electrodes: Comparison of mean-field approaches}

\author{Yuri~Yu.~Tarasevich}
\email[Corresponding author: ]{tarasevich@asu.edu.ru}

\author{Andrei~V.~Eserkepov}
\email{dantealigjery49@gmail.com}

\author{Irina~V.~Vodolazskaya}
\email{vodolazskaya\_agu@mail.ru}

\affiliation{Laboratory of Mathematical Modeling, Astrakhan State University, Astrakhan 414056, Russia}

\date{\today}

\begin{abstract}
We mimic nanorod-based transparent electrodes as random resistor networks (RRN) produced by the homogeneous, isotropic, and random deposition of conductive zero-width sticks onto an insulating substrate. We suppose that the number density (the number of objects per unit area of the surface) of these sticks exceeds the percolation threshold, i.e., the system under consideration is a conductor. We computed the electrical conductivity of random resistor networks vs the number density of conductive fillers for the wire-resistance-dominated case, for the junction-resistance-dominated case, and for an intermediate case. We also offer a consistent continuous variant of the mean-field approach. The results of the RRN computations were compared with this mean-field approach. Our computations  suggest that, for a qualitative description of the behavior of the electrical conductivity in relation to the number density of conductive wires, the mean-field approximation can be successfully applied when the number density of the fillers  $n > 2n_c$, where  $n_c$ is the percolation threshold. However, note the mean-field approach slightly overestimates the electrical conductivity. We demonstrate that this overestimate is caused by the junction potential distribution.
\end{abstract}

\maketitle

\section{Introduction\label{sec:intro}}

Transparent electrodes are important components of modern optoelectronic devices such as  touch-screens, heaters, and solar cells~\cite{Hecht2011AM,McCoul2016AEM,Sannicolo2016Small,Gao2016AdvPhys,ZhangChemRev2020,Patil2020}. Numerous efforts are currently underway to identify the main factors affecting the effective electrical conductivity, transparency, and haze of such films~\cite{Wang2008MSE,Mutiso2015PPS}. To characterize both the sheet resistance, $R_\Box$, and the transparency, $T$, of transparent electrodes, different figures of merit (FoMs) are  used~\cite{Fraser1972JES,Haacke1976JAP,Hecht2011AM,De2009ACSN,Han2018}.

One of the most widely used kinds of transparent electrode consists of a transparent, poorly conductive film containing randomly distributed highly conductive elongated fillers such as nanowires, nanotubes, or  nanorods~\cite{Nam2016N,Ackermann2016SR,Callaghan2016PCCP,McCoul2016AEM,Zhang2017JMM,Hicks2018JAP,Glier2020,Balberg2020}. Transparent electrodes should simultaneously have both high conductivity and high transparency. However, high transparency and high conductivity are mutually exclusive properties, since high transparency requires a low concentration of conductive fillers, while high conductivity implies their high concentration.

The simplest consideration suggests that the transmittance of a film is proportional to the expected fraction of its surface not covered by randomly deposited opaque objects
\begin{equation}\label{eq:T}
T = T_0 \mathrm{e}^{- n  a},
\end{equation}
where each object has an area $a$ (see, e.g., Ref.~\onlinecite{Yi2004JAP}). Here, $n$ is the number density, i.e., the number of objects, $N$, per unit area of the surface, $A$,
\begin{equation}\label{eq:numdensity}
  n = \frac{N}{A}.
\end{equation}
At low coverage ($n a \ll 1$), \eqref{eq:T} can be written as
\begin{equation}\label{eq:Tlowcoverage}
T \approx T_0 (1- n  a).
\end{equation}
Such linear dependency of the transparency on the surface coverage is consistent with experimental data~\cite{Bergin2012Nanoscale,Khanarian2013JAP}.

There are two main approaches to describe the electrical conductivity of random nanowire networks, viz., an exact one, based on the solution of huge systems of linear equations from Kirchhoff's rules and Ohm's laws (see, e.g., Refs.~\onlinecite{Kim2018JAP,Kim2020JCPC}), and a mean-field approach (MFA) (see, e.g., Refs.~\onlinecite{Kumar2017JAP,Forro2018ACSN}). Rather than study all the conductors in a system, an MFA involves considering a single conductor, placed in the mean  electric field that is produced by all the other conductors. However, even within an MFA, different variants can be used.

Thus, Ref.~\onlinecite{Kumar2017JAP} consistently uses a discrete consideration, viz., the contacts on the wire are discretely arranged, while the electrical current between any pair of nearest contacts is constant. By contrast, Ref.~\onlinecite{Forro2018ACSN} uses a hybrid discrete-continuous approach, viz., the contacts on the wire are discretely arranged, while the electrical current in the whole wire changes continuously except for the two end segments where  electrical currents are absent. (See Supplemental Material for a sketch of the current distribution in a conductive wire in different variants of the mean-field approach.)

Based on a geometrical consideration of a thin film of randomly deposited conductive wires, a formula for the sheet resistance has been proposed~\cite{Kumar2017JAP}
\begin{equation}\label{eq:KumarR}
  R_\Box = \frac{\pi}{2\sqrt{N_E}} \left( \frac{4\rho}{\pi D^2} + \frac{R_\text{j}}{d}\right),
\end{equation}
where $\rho$ is the electrical resistivity of the wire material, $D$ is the wire diameter, while the wire length, $l$, is assumed to be unity, $R_\text{j}$ is the junction resistance, and $N_E = n[Cn + \exp(-Cn) - 1]$ is the number density of wire segments,
$$
d = \frac{1 - \exp(-Cn)}{ Cn} - \exp(-Cn)
$$
is the mean segment length, and $C = 2/\pi$. Here, the number density of the conductive wires, $n$, is supposed to be high. The sheet resistance can be rewritten as follows
$$
R_\Box = \frac{\pi}{2 d \sqrt{N_E}} \left( R_n + R_\text{j} \right)=\\ \frac{\pi}{2 d \sqrt{N_E}} \left( d R_\text{s} + R_\text{j} \right),
$$
where $R_n$ is the averaged value of the electrical resistance between two junctions, and $R_\text{s}$ is the wire resistance.

An alternative formula has been proposed in Ref.~\onlinecite{Forro2018ACSN}.
\begin{multline}\label{eq:Forro}
  R_\Box  = \frac{R_\text{s}}{n l^2}\times\\ \left[ \frac{r_\text{m}}{2} - \sqrt{ \frac{R_\text{j} r_\text{m}}{ R_\text{s} C n l^2}}\tanh \left(\sqrt{\frac{R_\text{s} C n l^2 r_\text{m}}{4 R_\text{j}}}\right)\right]^{-1},
\end{multline}
where
$$
r_\text{m} = \frac{n_\text{a} - 1 +R_\text{j}\left( R_\text{j} + \frac{R_\text{s}}{n_\text{a} + 1} \right)^{-1}}{n_\text{a} + 1}, \quad n_\text{a} = nCl^2/2.
$$
(We have changed the original notation to provide uniformity throughout this text.)

The effect of junction resistance on the conductivity of nanowire- and nanotube-based conductive networks has been analyzed~\cite{Zezelj2012,Rocha2015,Ponzoni2019,Fata2020JAP}. Typically, in untreated nanowire-based networks, the wire-to-wire junction resistances dominate over the resistance of the nanowires, themselves (see, e.g., Refs.~\onlinecite{Bellew2015,Manning2020}).

Poorly conductive films containing randomly distributed highly conductive, elongated fillers are  kinds of inhomogeneous media. For several decades, the physical properties of inhomogeneous media have aroused the interest of the scientific community~\cite{McLachlan2007JNM}. Most attention has been  paid to the electrical properties of binary materials. There are different theories and models relating to the electrical conductivity of mixtures of conducting and insulating species. The effective medium approximation~\cite{Bruggeman1935AnnPhys} provides a good description of the physical properties at any concentration except for the fairly narrow region around the percolation threshold. The so-called generalized effective medium equation accounts for the position of the percolation threshold and the values of the conductivity exponents below and above percolation~\cite{McLachlan2007JNM}. Another tool used to describe the composites is percolation theory~\cite{Stauffer}.

The goal of the present work is an investigation of the electrical properties of 2D disordered systems with an insulating host matrix and conductive rod-like fillers (zero-width sticks). The number density of the conductive fillers ranges from the percolation threshold, $n_\text{c}$, to $\approx 20 n_\text{c}$. The system under consideration is treated as a random resistor network (RRN). The potentials and currents in any RRN can be found using Ohm's law and Kirchhoff's rules~\cite{Kirkpatrick1971PRL,Kirkpatrick1973RMP,Li2007JPhA,Benda2019,Kim2020JCPC}. By contrast, we consider a particular kind of MFA. Here, our approach is consistently continuous, i.e., the contacts are continuously spread over the conductor; the electrical current changes continuously in each conductive wire. This consideration implies a very high concentration of conductors.
Three limiting cases represent our focus, viz.,
\begin{enumerate}
  \item Unwelded wires. The junction resistance dominates over the wire resistance ($R_\text{j} \gg R_\text{s}$).
  \item Welded wires (see Ref.~\onlinecite{Ding2020} for a review of welding techniques). The junction resistance and the wire resistance are of the same order (in our study, $R_\text{j} = R_\text{s}$).
  \item Speculative case. The wire resistance dominates over the junction resistance  ($R_\text{j} \ll R_\text{s}$).
\end{enumerate}

The rest of the paper is constructed as follows. Section~\ref{sec:methods} describes some technical details of the simulation and our variant of the MFA. Section~\ref{sec:results} presents our main findings. In Section~\ref{sec:discussion}, we discuss the reliability of the presented results. Section~\ref{sec:concl} summarizes the main results.

\section{Methods\label{sec:methods}}

\subsection{Sampling}\label{subsec:sampling}
The particular case of a plane graph involves a situation with $N$ zero-width sticks of length $l$, the centers of which are assumed to be independent and identically distributed (i.i.d.)  within a square domain $\mathcal{D}$ of size $L \times L$ with periodic boundary conditions; $\mathcal{D} \in \mathbb{R}^2$, i.e., $x,y \in [0;L]$, where $(x,y)$ are the coordinates of the center of the stick under consideration. Their orientations are assumed to be equiprobable. Hence, a homogeneous and isotropic network is produced. The relation $L>l$ is assumed. In our simulations, without loss of generality, sticks of unit length were used ($l = 1$). These sticks were randomly deposited onto $\mathcal{D}$  until the desired number density was reached. For basic computations, we used a  system of size $L=32$. Each stick was treated as a resistor with a specified electrical conductivity, $\sigma_\text{s}$, i.e., an  RRN was considered.

To detect the percolation cluster, the Union--Find algorithm~\cite{Newman2000PRL,Newman2001PRE} modified for continuous systems~\cite{Li2009PRE,Mertens2012PRE} was applied. When a percolation  cluster was found, all other clusters were removed since they cannot contribute to the electrical conductivity. An adjacency matrix was formed for the percolation cluster. With this adjacency matrix in hand, Kirchhoff's current law was used for each junction of the sticks, and Ohm’s law for each circuit between any two junctions. The obtained set of linear equations with a sparse matrix has been solved using \emph{Eigen}~\cite{eigenweb}, a C++ template library for linear algebra. Since only square samples were considered, the electrical conductivity is simply the inverse of the sheet resistance, i.e., $\sigma = R_\Box^{-1}$.

The computer experiments were repeated 100 times for each value of the number density. The error bars in the figures correspond to the standard deviation of the mean. When not shown explicitly, they are of the order of the marker size.

\subsection{Mean-field approximation}\label{subsec:MFA}
\begin{figure}[htb]
  \centering
  \includegraphics[width=0.9\columnwidth]{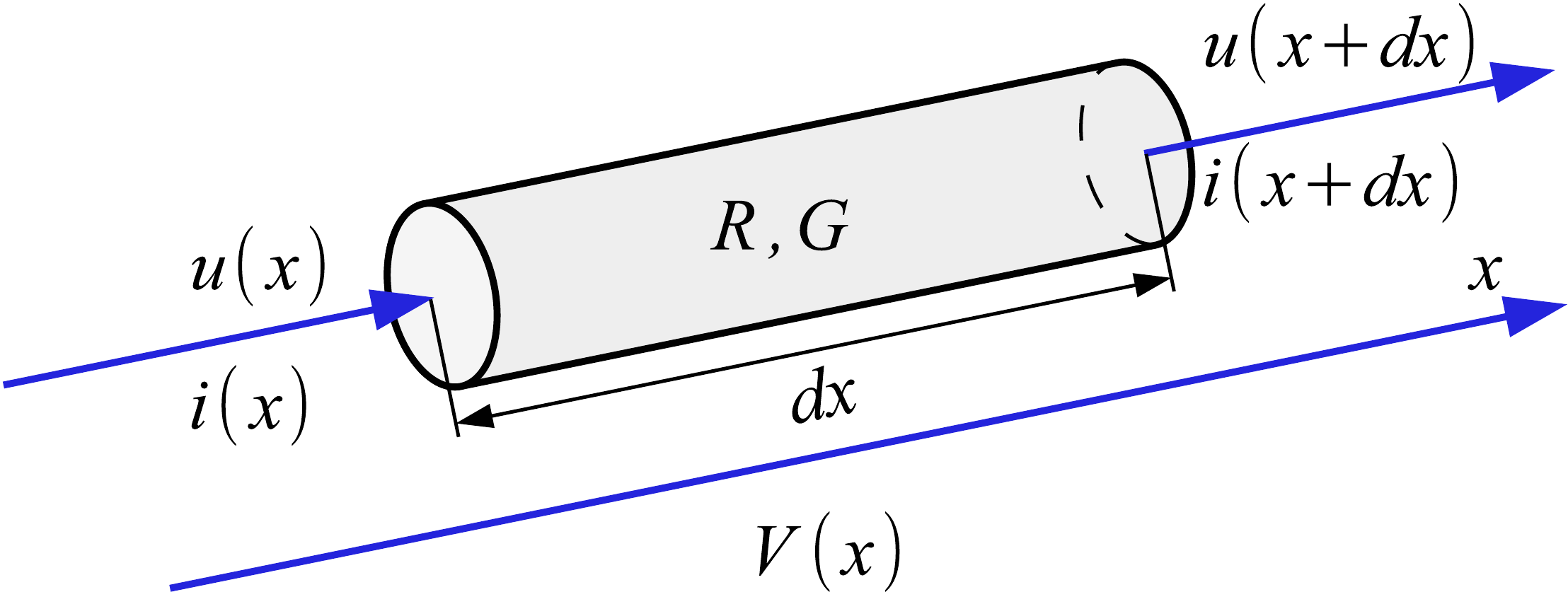}
  \caption{The conductive linear wire is placed in an external electrical field with a coordinate-dependent potential $V (x)$. The wire is characterized by its resistance per unit length, $R$, and its leakage conductivity per unit length, $G$.}\label{fig:wire}
\end{figure}

Let there be a linear conductive wire, the lateral surface of which is covered by an isolator. The conductive wire is characterized by the resistance, $R$, while the isolator is characterized by the leakage conductivity, $G$. Both quantities are referred to the unit length of the conductive wire. This wire is placed in an external electrical field with a coordinate-dependent potential $V (x)$. Consider a line segment of the wire, located between points of which the coordinates are $x$ and $x + dx$ (Fig.~\ref{fig:wire}).

In this line segment, the potential difference is $ u(x + dx) -  u(x) = \frac{d  u (x)}{d  x} dx$, while the voltage drop is $i(x) R d x$. According to Ohm's law,
\begin{equation}\label{eq:dudx}
\frac{d  u(x)}{d  x} + i(x) R  = 0,
\end{equation}
since no electromotive force acts in this line segment. The change in the electrical current in this line segment $ i(x + dx) -  i(x) = \frac{d  i (x)}{d  x} d x$ is due to a loss of charge caused by the imperfect isolator $[u(x) - V(x)] G dx$. Hence,
\begin{equation}\label{eq:didx}
\frac{d i (x)}{d x}  + [u(x) -V(x)] G = 0.
\end{equation}
The set of ordinary differential equations (ODEs)~\eqref{eq:dudx} and~\eqref{eq:didx}  can be treated as a type of stationary telegraph equations.

Differentiating ODE~\eqref{eq:dudx} with respect to $x$ and substituting ODE~\eqref{eq:didx} into ODE~\eqref{eq:dudx}, one can obtain an inhomogeneous linear ODE of the second order
\begin{equation}\label{eq:inhomod2udx2}
\frac{d^2 u(x)}{d x^2} - \lambda^2 u(x) = -\lambda^2 V(x).
\end{equation}
where $\lambda = \sqrt{RG}$. The solution to this ODE is equal to the sum of the general solution of the homogeneous ODE
$$
u_0(x) = A_1 \exp(\lambda x) + A_2 \exp(-\lambda x)
$$
and a particular solution of the inhomogeneous ODE $u^\ast(x)$ that  depends on the particular type  of the function $V(x)$.

Let a linear wire of length $l$ be located at an angle $\alpha$ to a uniform electric field $E = V_0/L$. The potential of this field is $V(x) = - x E \cos \alpha$, when the axis $x$ is directed along the wire, while the axis origin coincides with the wire center. For this particular sort of $V(x)$, $u^\ast(x) = - x E \cos \alpha $.
Thus,
$$
u(x) = A_1 \exp(\lambda x) + A_2 \exp(-\lambda x) - x E \cos \alpha.
$$
Since
\begin{equation}\label{eq:i}
i(x) = - \frac{1}{R}\frac{d u(x)}{d x},
\end{equation}
according to~\eqref{eq:dudx},
\begin{equation}\label{eq:i1}
i(x) = \frac{E}{R} \cos \alpha - \frac{\lambda}{R} \left[ A_1 \exp(\lambda x) - A_2 \exp(-\lambda x)\right].
\end{equation}
In our consideration, the leakage current is associated exclusively with the lateral surface of the wire, therefore there should be no electrical current at the ends of the wire, i.e., $i(-l/2)=i(l/2)=0$. Thus,
\begin{equation}\label{eq:currentMFA}
i(x; \lambda, \alpha) = \frac{E}{R} \left[ 1 - \frac{\cosh(\lambda x)}{\cosh\left(\frac{\lambda l}{2}\right)}\right]\cos \alpha .
\end{equation}
The electrical current depends both on the coordinate $x$ and on the parameters $\lambda$ and $\alpha$. The electrical current averaged along the wire length is
$$
\langle i(\lambda, \alpha) \rangle = \frac{1}{l} \int_{-\frac{l}{2}}^{\frac{l}{2}} i(x; \lambda, \alpha) \, dx ,
$$
i.e.,
$$
\langle i(\lambda, \alpha) \rangle = \frac{E}{R} \left[ 1 - \frac{2}{l \lambda}  \tanh\left(\frac{\lambda l}{2}\right)\right]\cos \alpha.
$$

Now we can turn to considering the system described in Section~\ref{subsec:sampling}. When the system under consideration is dense ($n \gtrapprox 2 n_c$), the potential drop along the system is linear~\cite{Sannicolo2018,Forro2018ACSN,Papanastasiou2021,Charvin2021}. Instead of a consideration of the random resistor network produced by all the sticks, here, there is consideration of only one stick in the mean-field produced by all the other sticks. Since all orientations of a stick are equiprobable, the number of sticks intersecting a line of width $L$ perpendicular to the field is $ n l L \cos \alpha $. The total electrical current in the sticks of all orientations through a cross section of the system is
\begin{multline*}
\langle \mathcal{I}(\lambda) \rangle =\\ \frac{1}{\pi} \int_{-\pi/2}^{\pi/2} \frac{E n l L}{R} \left[ 1 - \frac{2}{l \lambda}  \tanh\left(\frac{\lambda l}{2}\right)\right] \cos^2 \alpha \, d \alpha =\\  \frac{E n l L}{2 R} \left[ 1 - \frac{2}{l \lambda}  \tanh\left(\frac{\lambda l}{2}\right)\right].
\end{multline*}

When the resistance of the stick is $R_\text{s}$, the resistance per the unit length is
$$
R = \frac{R_\text{s}}{l}.
$$
When the junction resistance between any two sticks is $R_\text{j}$, the leakage conductivity per unit length is
$$
G = \frac{k}{l R_\text{j}},
$$
where $k$ is the number of junctions between the given stick and other sticks.

The number of junctions obeys a binomial distribution, which transforms at high concentrations of conductors into the Poisson distribution  $k \sim \mathrm{Pois}(C n l^2),$ where $C = 2/\pi$.
The probability distribution function (PDF) is
$$
p(k) = (C n l^2)^k \frac{\mathrm{e}^{-C n l^2}}{k!} \approx \frac{1}{\sqrt{2 \pi C n l^2}} \exp \left[ -\frac{(k - C n l^2)^2}{2 C n l^2 } \right].
$$
When the values of $n$ are large, i.e., for the very case when the MFA is appropriate, the PDF is a narrow peak. Therefore, the average value of the term depending on the number of junctions can be calculated approximately by substituting the mathematical expectation of the number of junctions, i.e., $C n l^2$,
\begin{multline*}
\left\langle\frac{2}{\lambda_k l} \tanh\left(\frac{\lambda_k l}{2}\right) \right\rangle = \sum_k p(k)\frac{2}{\lambda_k l} \tanh\left(\frac{\lambda_k l}{2}\right)  \approx\\
\frac{2}{\left\langle\lambda_k \right\rangle l} \tanh\left(\frac{\left\langle\lambda_k \right\rangle l}{2}\right),
\text{ where }
\lambda_k^2 =  \frac{k R_\text{s}}{l^2 R_\text{j}}.
\end{multline*}
Here, the subscript $k$ is intended to indicate the dependence of the parameter $\lambda$ on the number of inter-stick junctions.

Then, the electrical conductivity of the system under consideration is
\begin{equation}\label{eq:MFAsigma}
\sigma = \frac{  n l^2 }{2 R_\text{s}} \left[ 1 - \sqrt{\frac{4 R_\text{j}}{ n l^2 R_\text{s} C} } \tanh\left(\sqrt{\frac{ n l^2 R_\text{s} C}{4 R_\text{j}} }\right)\right].
\end{equation}
This formula is closely related to formula~\eqref{eq:Forro}, which can be rewritten as
\begin{multline}\label{eq:ForroCond}
  \sigma  = \frac{r_\text{m} n l^2}{2 R_\text{s}}\times\\ \left[ 1 - \sqrt{ \frac{4 R_\text{j}}{r_\text{m} n l^2 R_\text{s} C}}\tanh \left(\sqrt{\frac{r_\text{m} n l^2 R_\text{s} C }{4 R_\text{j}}}\right)\right].
\end{multline}
Formulae~\eqref{eq:MFAsigma} and~\eqref{eq:ForroCond} coincide up to the replacement $n \to r_\text{m} n$ where only one adjustable parameter $r_\text{m}$ (the so-called ``effective wire length'') is near to unity when $n \gg 1$. Both formulae are based on the assumption that all conductors contribute to the electrical conductivity, i.e., all conductors belong to the percolation cluster. This assumption is valid for $n \gtrapprox 1.5 n_\text{c}$~\cite{Kumar2017JAP,Tarasevich2021PREbb}.

\subsection{Figure of merit}\label{subsec:FoM}
FoM can be defined in different ways, e.g.,
\begin{equation}\label{eq:FoM1}
  \Phi_\text{TC} = \frac{T}{R_\Box}
\end{equation}
(see \onlinecite{Fraser1972JES} for details)
and
\begin{equation}\label{eq:FoM2}
  \Phi_\text{TC} = \frac{T^{10}}{R_\Box}
\end{equation}
(see~\cite{Haacke1976JAP} for details). An analysis of the advantages and disadvantages of each of these FoMs  can be found in Ref.~\onlinecite{Cisneroscontreras2019RP}. Another FoM has often been  used over the past few years (see, e.g., Refs.~\onlinecite{Hecht2011AM,De2009ACSN,Han2018})
\begin{equation}\label{eq:FoM3}
  \Phi_\text{TC} = \frac{\sigma_\text{DC}}{\sigma_\text{opt}(\lambda)},
\end{equation}
where the transparency, $T(\lambda)$, the optical conductance, $\sigma_\text{opt}(\lambda)$, the DC film conductance, $\sigma_\text{DC}$, and the sheet resistance,  $R_\Box$, are connected as follows
\begin{equation}\label{eq:Tlambda}
T(\lambda) =  \left[1 + \frac{1}{2R_\Box} \sqrt{\frac{\mu_0}{\varepsilon_0}} \frac{\sigma_\text{opt}(\lambda)}{\sigma_\text{DC}}  \right]^{-2}.
\end{equation}
Here, $\varepsilon_0$ and $\mu_0$ are the electrical and magnetic constants, respectively, and $\lambda$ is the wave length. In this way, Eq.~\eqref{eq:FoM3} transforms into
\begin{equation}\label{eq:FoM3final}
  \Phi_\text{TC} = \frac{188.5}{R_\Box}\frac{\sqrt{T(\lambda)}}{1-\sqrt{T(\lambda)}}.
\end{equation}

For typical nanorods with aspect ratios up to $10^3$ (see, e.g., Refs.~\onlinecite{Vodolazskaya2019JAP,Forro2018ACSN}) deposited onto a transparent substrate with number densities of up to $\approx 10n_\text{c}$, the linear dependency of the transmittance on the number density of the deposited nanorods~\eqref{eq:Tlowcoverage} is not only valid, but can even be simplified to $T\approx T_0$. Hence, the FoMs~\eqref{eq:FoM1} and~\eqref{eq:FoM2} can be considered to be approximately inversely proportional to the sheet resistance
$$
\Phi_\text{TC} = \frac{\Phi_0}{R_\Box},
$$
where $\Phi_0$ is a constant.

\section{Results\label{sec:results}}

\begin{figure}[!htb]
\includegraphics[width=\columnwidth]{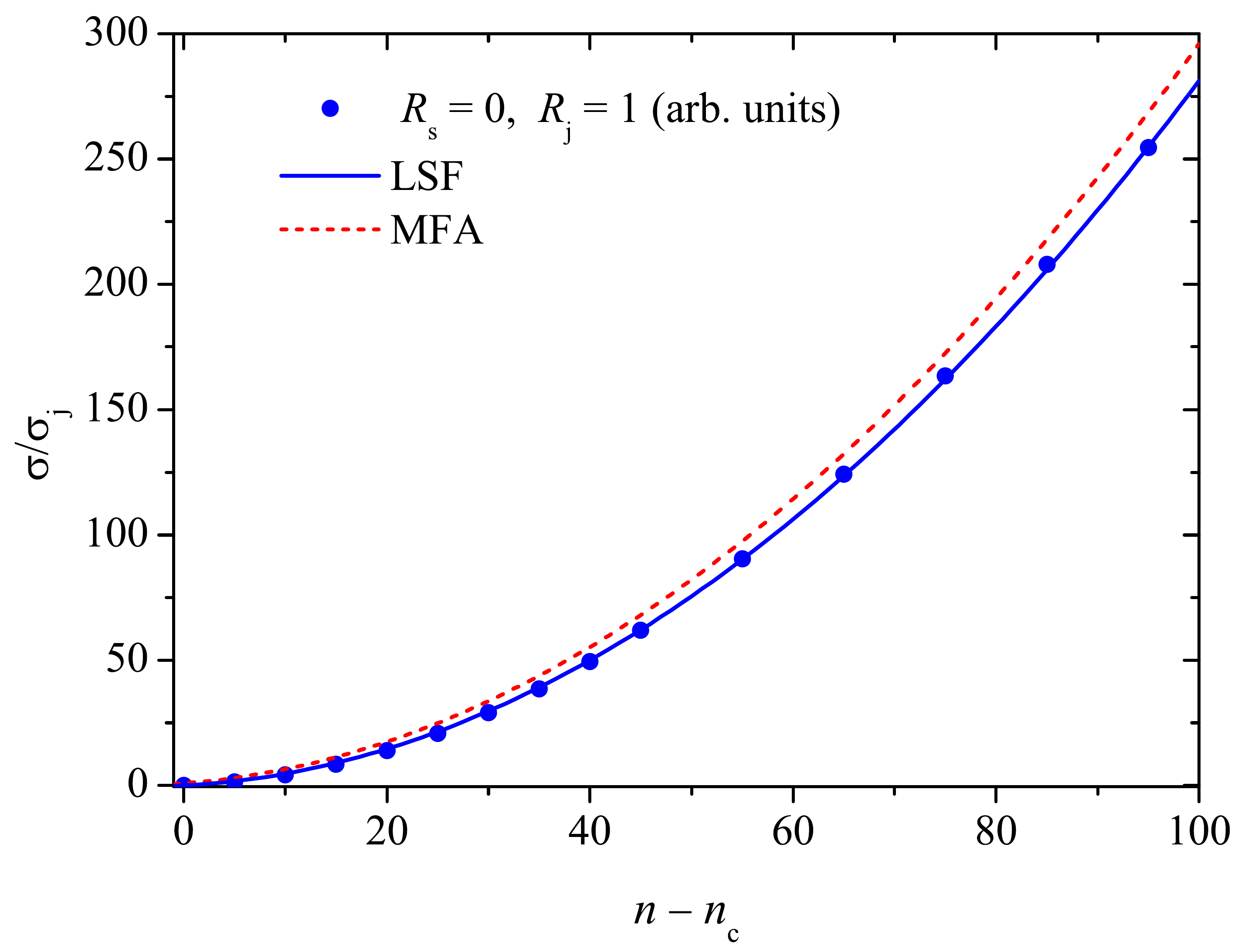}\\
\caption{Comparison of the dependencies of the electrical conductivity, $\sigma$, on the number density, $n$, for $L=32$, $R_\text{s}=0$, $R_\text{j}=1$ arb. units. Solid symbols correspond to our simulation results. The solid line corresponds to the least squares fit $\sigma/\sigma_\text{j} = 0.019 + 0.21(n - n_\text{c}) + 0.026(n - n_\text{c})^2$.  The dashed curve corresponds to the MFA~\eqref{eq:MFAsigma}.\label{fig:conductivityJDR}}
\end{figure}

Figure~\ref{fig:conductivityJDR} shows the dependency of the electrical conductivity, $\sigma$, on the number density for the junction-resistance-dominated case ($R_\text{j} \gg R_\text{s}$). The electrical conductivity can be fitted by the second order polynomial  $\sigma/\sigma_\text{j}  = 0.019 + 0.21(n - n_\text{c}) + 0.026(n - n_\text{c})^2$.
When $n \gg 1$, formula~\eqref{eq:MFAsigma} simplifies to
\begin{equation}\label{eq:MFA-JDR}
\sigma = \frac{ C  n^2 l^4 }{24 R_j},
\end{equation}
i.e., $\sigma \approx 0.265 n^2$, which is fairly consistent with the direct, RRN, computations of the electrical conductivity.

When $n \gg 1$, formula~\eqref{eq:KumarR} simplifies to
\begin{equation}\label{eq:KumarJDR}
\sigma = \frac{\sqrt{C}}{R_\text{j} }  \approx \frac{0.8}{R_\text{j}},
\end{equation}
i.e., the predicted asymptotic behavior of the electrical conductivity radically differs from the results of the direct computations.
For this limiting case, formula~\eqref{eq:Forro} (see Ref.~\onlinecite{Forro2018ACSN}) predicts
$$
\sigma = \frac{Cn^2}{24 R_\text{j} } \left( \frac{Cn}{Cn + 2}\right)^2.
$$
When $n \gg 1$, this formula simplifies to $\sigma \approx 0.0265 n^2/R_\text{j}$.

\begin{figure}[!htb]
\includegraphics[width=\columnwidth]{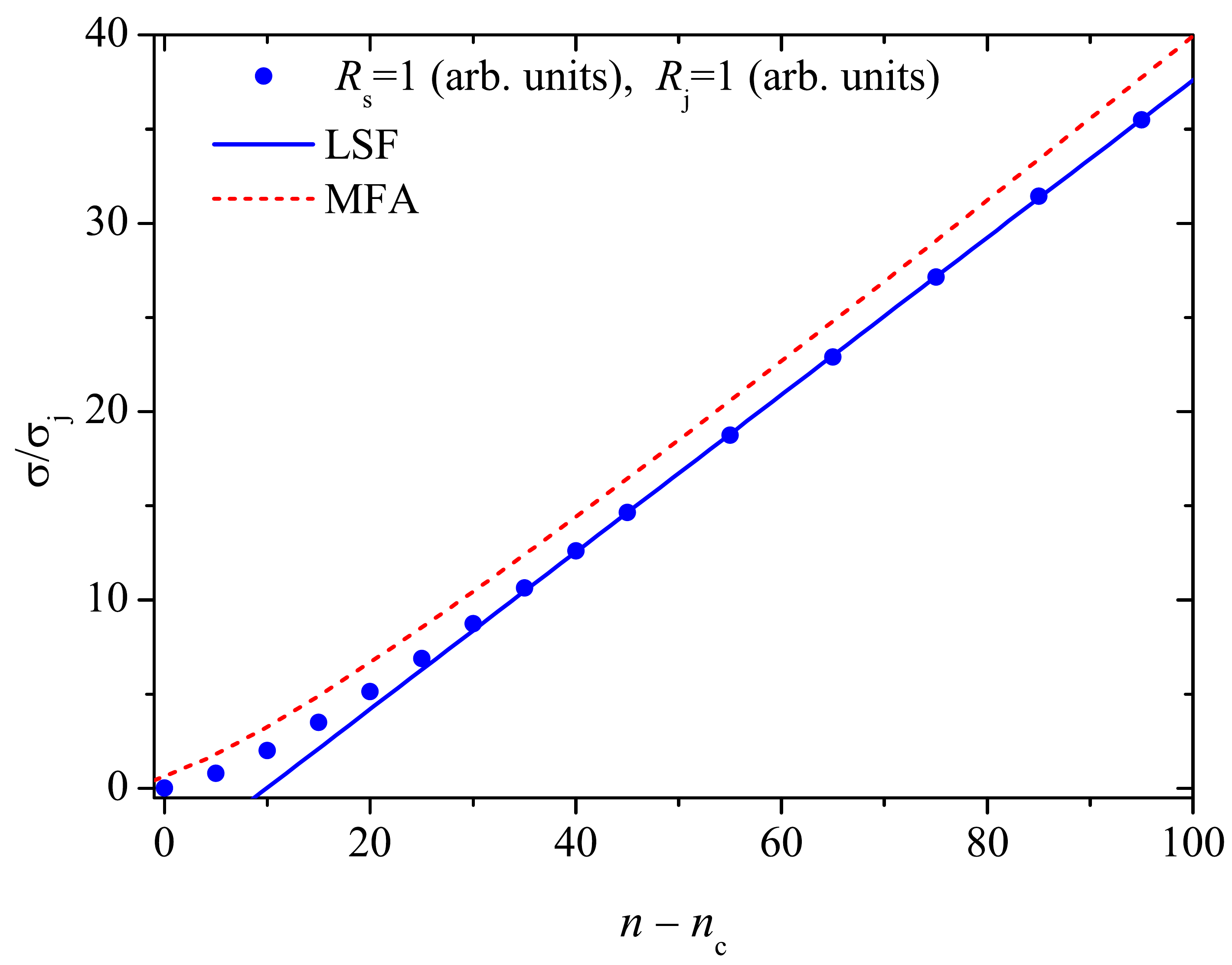}\\
\caption{Comparison of the dependencies of the electrical conductivity, $\sigma$, on the number density, $n$, for $L=32$, $R_\text{s}=1$ arb. units, $R_\text{j}=1$ arb. units. Solid symbols
correspond to our simulation results. The solid line corresponds to the least squares fit $\sigma/\sigma_\text{j} = -4.15 + 0.417(n - n_\text{c})$.  The dashed curve corresponds to formula~\eqref{eq:MFAsigma}.
\label{fig:conductivityJWR}}
\end{figure}

Figure~\ref{fig:conductivityJWR} shows the dependency of the electrical conductivity, $\sigma$, on the number density for the case when both the wire resistances and the junction resistances are equally important  ($R_\text{j} = R_\text{s}$ = 1 arb. units). The electrical conductivity can be fitted by the line $\sigma/\sigma_\text{j} = -4.15 + 0.417(n - n_\text{c})$.
When $n \gg 1$, the MFA prediction~\eqref{eq:MFAsigma} tends to
\begin{equation}\label{eq:MFA-JWR}
   \sigma = \frac{n l^2 }{2 },
\end{equation}
which is fairly consistent with the direct computations of the electrical conductivity. Formula~\eqref{eq:Forro} (see Ref.~\onlinecite{Forro2018ACSN}) demonstrates similar limiting behavior. By contrast,  formula~\eqref{eq:KumarR} (Ref.~\onlinecite{Kumar2017JAP}) tends to the limit value~\eqref{eq:KumarJDR} when $n \gg 1$.

\begin{figure}[!htb]
\includegraphics[width=\columnwidth]{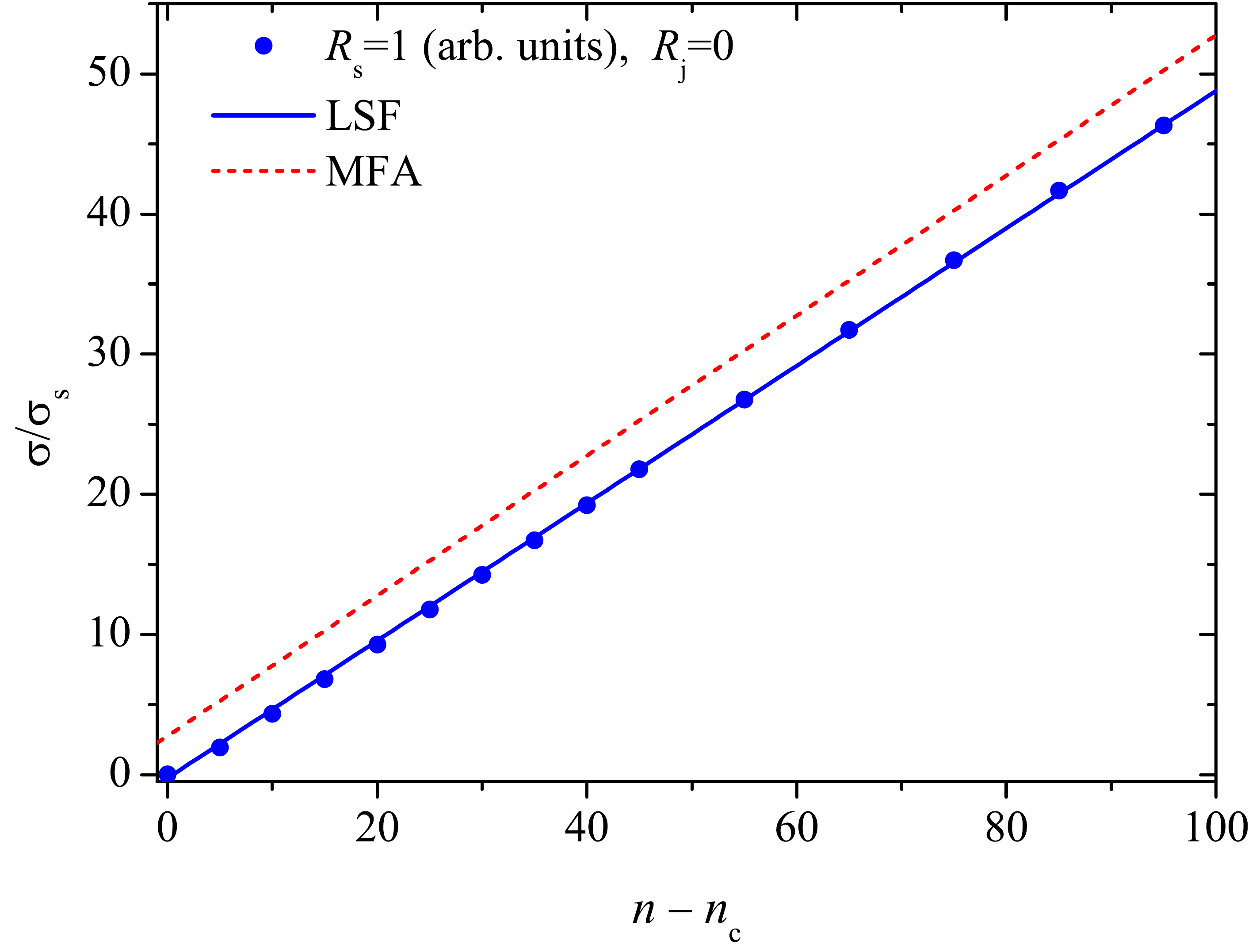}\\
\caption{Comparison of the dependencies of the electrical conductivity, $\sigma$, on the number density, $n$, for $L=32$, $R_\text{s}=1$ arb. units, $R_\text{j}=0$. Solid symbols correspond to our simulation results. The solid line corresponds to the linear least squares fit $\sigma/\sigma_\text{s} = - 0.26 + 0.49(n-n_\text{c})$. The dashed curve corresponds to formula~\eqref{eq:MFAsigma}.
\label{fig:conductivity}}
\end{figure}

Figure~\ref{fig:conductivity} shows the dependency of the electrical conductivity, $\sigma$, on the number density for a wire-resistance-dominated  case ($R_\text{s} \gg R_\text{j}$). When $n \gtrapprox 2n_\text{c}$, the electrical conductivity can be fitted by the linear function  $\sigma/\sigma_\text{s} = - 0.65 + 0.497(n - n_\text{c})$.
When $n \gg 1$, formula~\eqref{eq:KumarR} simplifies to
$$
\sigma = \frac{C^{3/2} n}{R_\text{s}},
$$
i.e., in our computation, $\sigma \approx 0.508 n$, since $R_\text{s}=1$.
For the wire-resistance-dominated case, formula~\eqref{eq:Forro} simplifies to
$$
\sigma = \frac{n}{2R_\text{s}}\frac{Cn-2}{Cn+2}.
$$
When $n \gg 1$, the asymptotic behavior of the electrical conductivity is
$$
\sigma \approx \frac{n}{2R_\text{s}},
$$
i.e., in our computation, $\sigma \approx 0.5n$, since $R_\text{s}=1$. Our simulation results are comparable to our MFA predictions Eq.~\eqref{eq:MFAsigma}, as well with the predictions of other authors~\eqref{eq:KumarR} (see Ref.~\onlinecite{Kumar2017JAP}) and~\eqref{eq:Forro} (see Ref.~\onlinecite{Forro2018ACSN}). Although the dependencies are presented by almost parallel lines, formula~\eqref{eq:MFAsigma} overestimates the electrical conductivity.

The significant difference between our simulation and the theoretical predictions (both~\eqref{eq:MFAsigma} and~\eqref{eq:KumarR} (see Ref.~\onlinecite{Kumar2017JAP})) may reflect a more complex network structure than that supposed in Ref.~\onlinecite{Kumar2017JAP}.
The use of an adjusted parameter~\cite{Forro2018ACSN} leads to behavior closer to our simulations in both the limiting cases, however the adjusted parameter has hardly any clear theoretical background. The differences between the MFA predictions and the direct computations of the electrical conductivity are discussed in  Section~\ref{sec:discussion}.

A simple estimate of the FoM can be performed for the wire-resistance-dominated case. When the aspect ratio (length-to-width ratio) of conductive rods is $\varepsilon$, the maximum of the FoM corresponds to the number density $n \approx \varepsilon$ for formula~\eqref{eq:FoM1} while $n \approx \varepsilon/10$ for formula~\eqref{eq:FoM2}.  Formula~\eqref{eq:FoM3} predicts the maximums of the FoMs $n \approx 53$ (for $\varepsilon = 100$) and $n \approx 163$ (for $\varepsilon = 1000$).

\section{Discussion\label{sec:discussion}}
Section~\ref{sec:results} evidenced that the MFA overestimates the electrical conductivity. (Despite some reasoning and computations~\cite{Forro2018ACSN} intended to justify the adjustable parameter  $r_\text{m}$ in formula~\eqref{eq:Forro}, this parameter can hardly be accepted as rigorously introduced. Omitting this rather unclear parameter, our approach and formula~\eqref{eq:Forro} are exactly identical.) Discrepancies between direct computations of the electrical conductivity and its evaluation within the MFA may be due to several reasons:
\begin{enumerate}
  \item Inaccurate evaluation of the number of wires that contribute to the electrical conductivity.
  \item Finite-size effect.
  \item Inaccurate evaluation of the mean values.
  \item Nonlinear variation of the electrical potential along the samples.
  \item Difference between the average electrical current in a stick obtained within the MFA and the actual average electrical current in the real system.
\end{enumerate}

Our evaluation is based on the assumption that all conductive sticks contribute to the electrical conductivity, i.e., all sticks belong to the percolation cluster, and that this does not contain dead ends other than the ends of the sticks. Two independent studies~\cite{Kumar2017JAP,Tarasevich2021PREbb} completely confirm this assumption. Namely, even when $n \approx 2n_\text{c}$, only a negligible fraction of the sticks does not belong to the percolation cluster. Moreover, the percolation cluster is identical to its geometrical backbone, excepting all the stick ends. Hence, the observed overestimation in the electrical conductivity may be due to significant differences between the geometrical backbone and the current-carrying part of the percolation cluster. This possibility, although it looks rather unlikely, nonetheless, needs to be further checked.

\begin{figure}[!htb]
\includegraphics[width=\columnwidth]{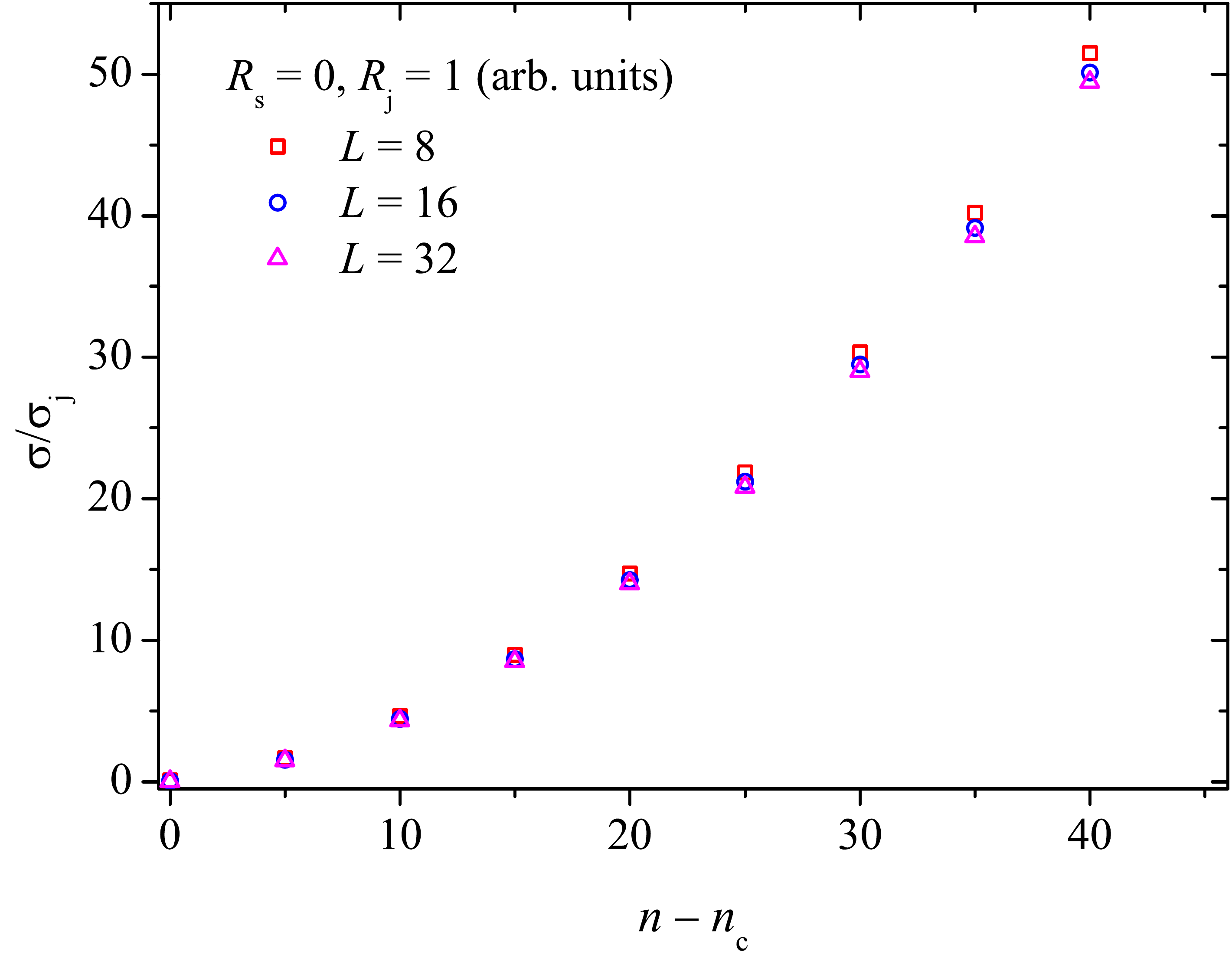}\\
\caption{Finite-size effect for a junction-resistance-dominated case.\label{fig:FSE-JDR}}
\end{figure}

Figure~\ref{fig:FSE-JDR} demonstrates the finite-size effect for a junction-resistance-dominated case. The system size shown was $L=8,16,32$. For the two other cases tested, the differences in the electrical conductivity, for systems of different sizes, were within the marker size, i.e., smaller than the statistical error (see Supplemental Material).

\begin{figure}[!htb]
\includegraphics[width=\columnwidth]{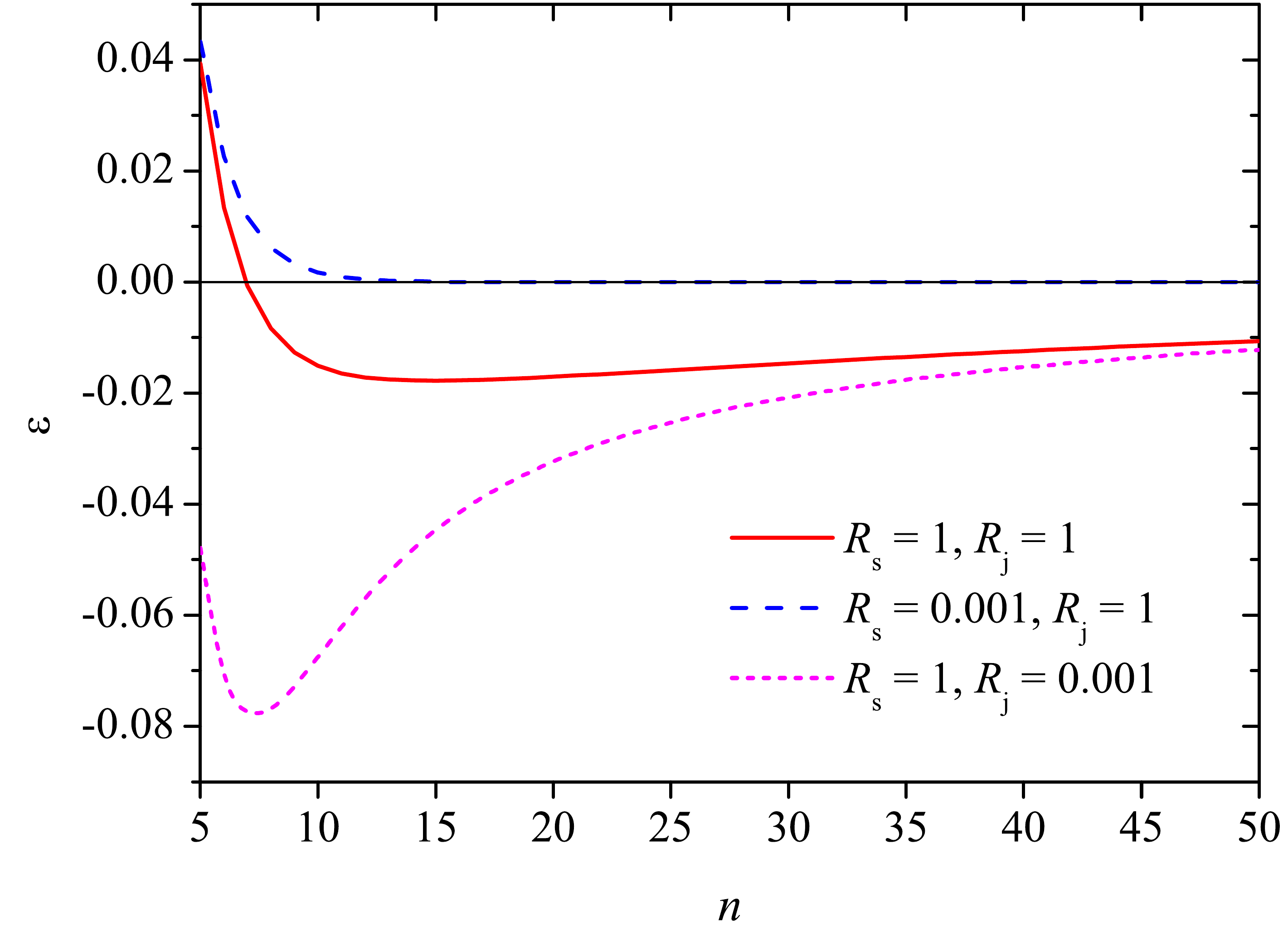}\\
\caption{Relative deviation between $\frac{2}{\left\langle\lambda_k \right\rangle l} \tanh\left(\frac{\left\langle\lambda_k \right\rangle l}{2}\right)$ and $\left\langle\frac{2}{\lambda_k l} \tanh\left(\frac{\lambda_k l}{2}\right) \right\rangle$ for different ratios between the stick resistance and junction resistance. \label{fig:deviation}}
\end{figure}

Figure~\ref{fig:deviation} evidenced that the estimate used for the mean value is fairly reasonable for dense systems, i.e., for the very case when the MFA is assumed to be valid. A larger electrical contrast (ratio between the wire resistance and the junction resistance) does not change the relative error. The exact mean value $\left\langle\frac{2}{\lambda_k l} \tanh\left(\frac{\lambda_k l}{2}\right) \right\rangle$ has been found using Maple.

\begin{figure}[!htb]
  \centering
  \includegraphics[width=\columnwidth]{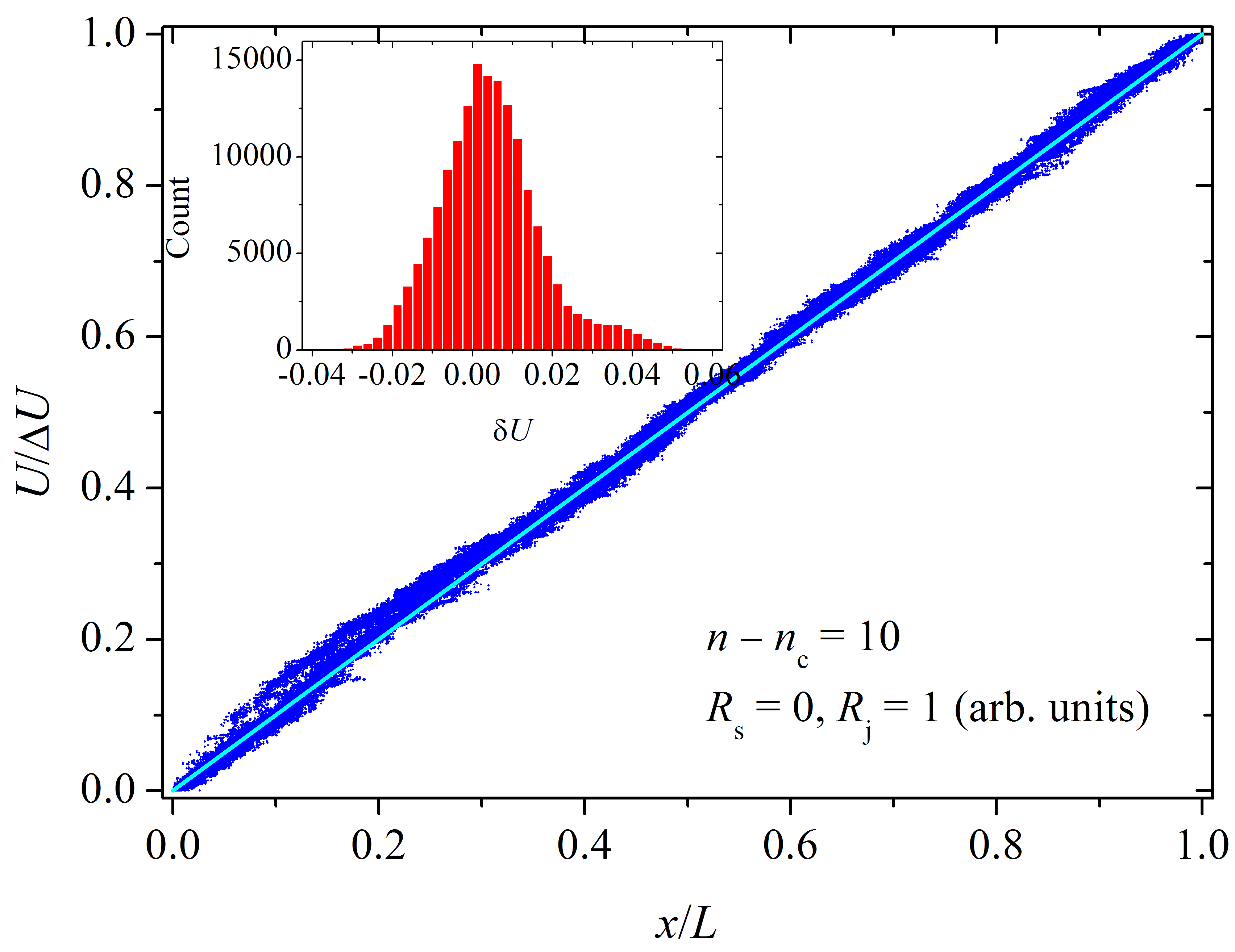}\\
  \caption{The normalized potential, $U/\Delta U$, of each junction of the network is plotted against its normalized position, $x/L$, in one particular sample at $n - n_\text{c}=10$ for the junction-resistance-dominated case. The potential difference, $\Delta U$, is applied along the $x$-axis. A line is presented for comparison. Inset: Distribution of the junction potentials relative to the line $U/\Delta U = x/L$.
  \label{fig:potential}}
\end{figure}

The linearity of the electrical potential in the samples can be confirmed by computation.  Figure~\ref{fig:potential} demonstrates one example of how the electrical potential depends on the position of the junctions. Additional examples (for other ratios between the wire resistance and junction resistance and for other filler concentrations) are presented in the Supplemental Material. Similar results has also been published elsewhere~\cite{Sannicolo2018,Forro2018ACSN,Tarasevich2019}.

\begin{figure}[!htb]
  \centering
  \includegraphics[width=\columnwidth]{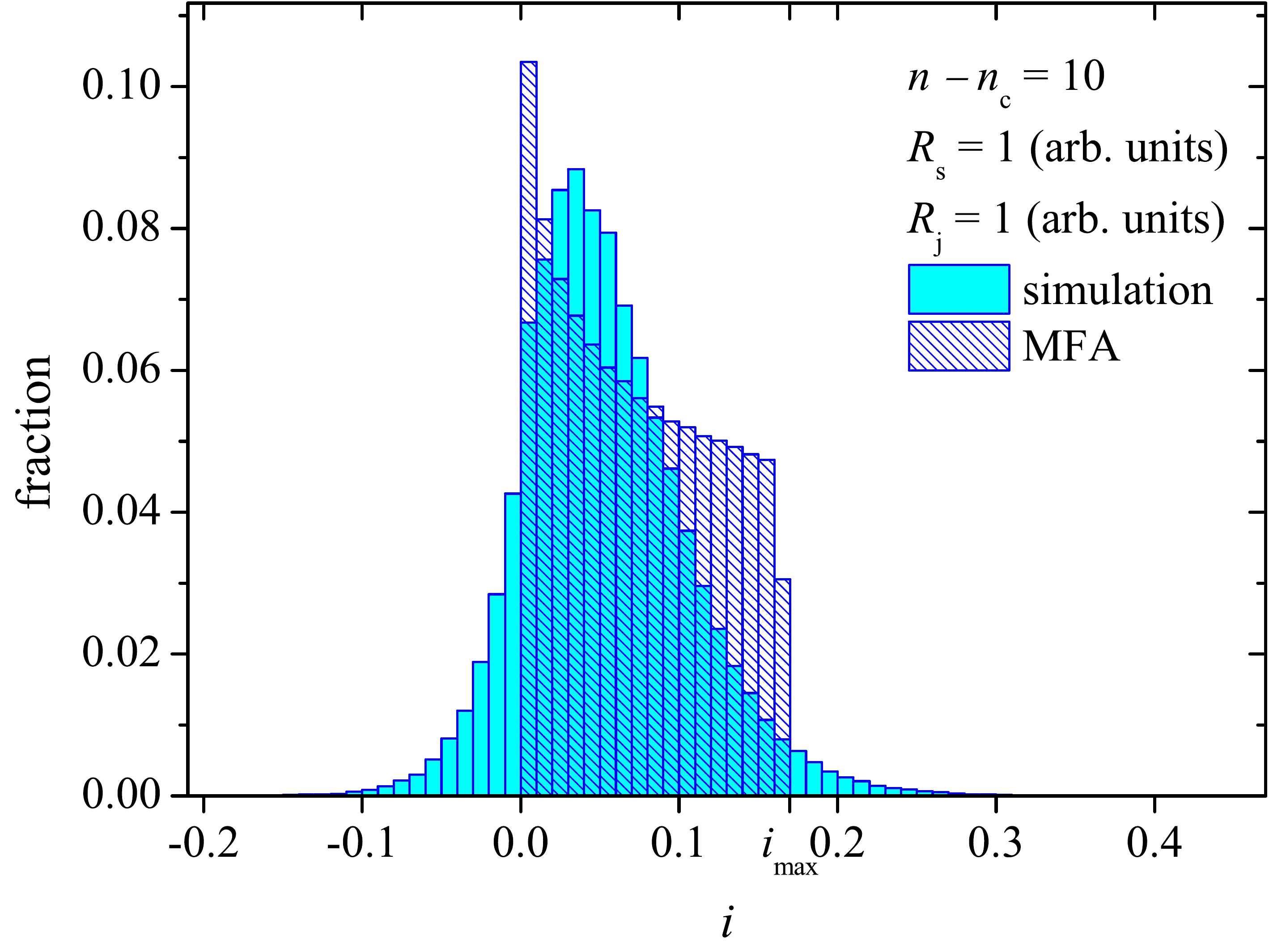}\\
  \caption{Distribution of the electrical current in a sample (solid fill) in comparison with the prediction of the mean-field approach (hatching) for the case when $R_\text{s} = R_\text{j} = 1$ (arb. units); $n - n_\text{c} = 10$. \label{fig:CurrentHist10}}
\end{figure}

The fact that the values of the potentials of the junctions are located not exactly on a straight line, but within a certain band, causes significant distortion of the current distribution in comparison with the predictions from the MFA. Figure~\ref{fig:CurrentHist10} demonstrates the current distribution in a sample for the case when $R_\text{s} = R_\text{j} = 1$; $n - n_\text{c} = 10$. Instead of the expected compact distribution with $i \in [0;i_\text{max}]$ predicted by the MFA (hatching), a diffuse distribution even with negative currents appears (solid fill). To generate the reference current distribution (MFA), the electrical current was calculated using formula~\eqref{eq:currentMFA} for $10^6$ independent random values of $x$ and $\alpha$.

\begin{figure}[!htb]
\includegraphics[width=\columnwidth]{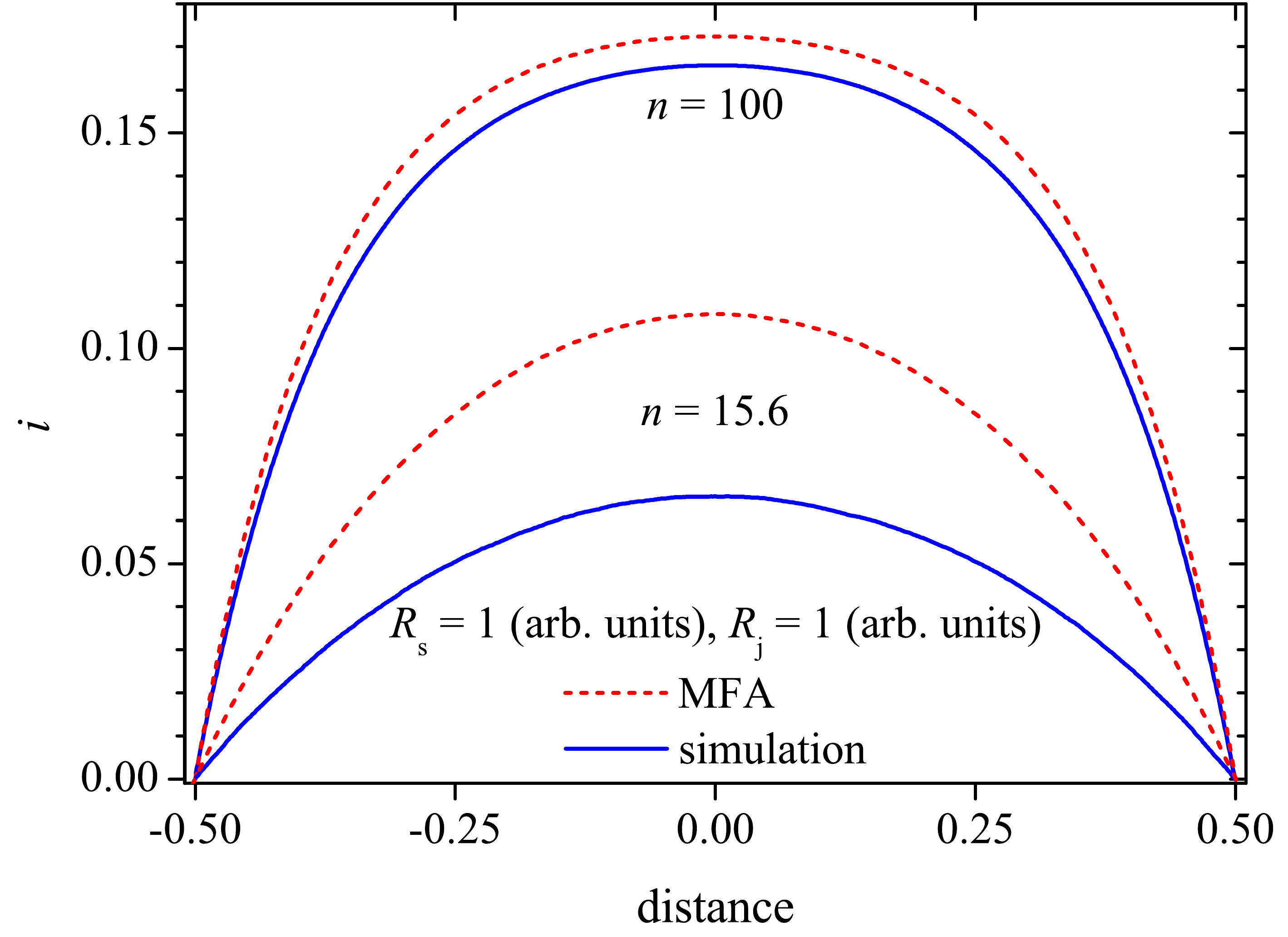}\\
\caption{Dependencies of the electrical current on the position in the conductive wire. The current is averaged over all the wires in one sample. The dashed curves correspond to the evaluation obtained within the mean-field approach while the solid curves correspond to the simulation. The wire resistance and the junction resistance are equal, $R_\text{s} = R_\text{j} = 1$ (arb. units);
$n = 15.6$ (two upper curves) and $n = 100$ (two bottom curves).
\label{fig:CurrentComparison10}}
\end{figure}

The difference between the actual distribution of the currents and the predictions of the MFA leads to a decrease in the mean electrical current in the single conductive wire as compared with the predictions of the MFA. Figure~\ref{fig:CurrentComparison10} demonstrates the dependencies of the mean electrical current against position in the conductive wire, for two values of the number density $n = 15.6$ and $n = 100$ when the wire resistance and the junction resistance are equal. The solid curves correspond to the evaluation obtained within the MFA while the dashed curves correspond to the simulation.

The MFA overestimates the electrical current. In the middle of the wire, the excess is about 50\% when $n = 15.6$. However, this effect decreases as the number density of the conductive wires increases. Thus, for $n=100$, the excess is only 5\%. This excess of the mean current corresponds to the excess of the electrical conductivity obtained within the MFA over the simulation results.

\section{Conclusion\label{sec:concl}}
We have considered random resistor networks produced by the homogeneous, isotropic, and random deposition of conductive sticks onto an insulating substrate. Using Kirchhoff's rules and Ohm's law, the electrical conductivity of such networks was calculated for a wide range of number densities of conductive wires from the percolation threshold up to $n = 100$. Moreover, the electrical properties of such networks have been studied using consistent continuous consideration within a mean-field approach. For different values of the wire resistance and junction resistance, the dependencies of the electrical conductivity on the number density of the conductive fillers have been obtained. Our study suggests that, for a qualitative description of the behavior of the electrical conductivity of random nanowire networks, the mean-field approach can be successfully applied when the concentration of fillers $n \gtrapprox 2n_c$. However, although the mean-field approach overestimates the electrical conductivity, the relative deviation decreases as the number density of the conductive wires increases, since our continuous consideration within the mean-field approach assumes a high number density of conductive wires. Estimates of the concentration of fillers corresponding to the optimum of the figure of merit indicate that the mean-field approximation is excellent and appropriate for practical purposes, since the optimum corresponds to the concentrations for which the theoretical estimates almost coincide with direct computer calculations based on Kirchhoff's rules.

Our study evidenced that the overestimation of the electrical conductivity within the mean-field approach as compared to the direct computations is due to changes of the electrical potential along the random resistor network not being not strictly linear. If the magnitude of the deviation from the linear law could be known \emph{a priori}, this would offer the possibility for introducing an appropriate correction into the consideration by the mean-field approach. However, at present, we can see no way for performing such an \emph{a priori} assessment. This is, probably, a promising topic for further research.

\section*{Supplementary material}

The supplementary material provides some additional information about the behavior of the electrical conductivity of the random resistor network under consideration. Figure S1 presents a sketch of the current distribution in a conductive wire in different variants of the mean-field approach.  Figure S2 demonstrates the absence of the finite-size effect for the wire-resistance-dominated case and in the case when both resistances are equal. Figure S3 presents the distribution of the electrical currents in segments of the wires. Additional examples of how the electrical potential depends on the position of the junctions are presented in Figure S4 (for other ratios between the wire resistance and junction resistance and for other filler concentrations).

\acknowledgments
Y.Y.T. and A.V.E. acknowledge the funding from the Foundation for the Advancement of Theoretical Physics and Mathematics ``BASIS'', grant~20-1-1-8-1.
The authors would also like to thank  A.G.Gorkun for technical assistance.

\bibliography{mfa}

\end{document}